\documentclass[12pt]{article}
\begin{document}

\author{A.I.Volokitin$^{1,2}$ and B.N.J.Persson$^1$ \\
%EndAName
\\
%EndAName
$^1$Institut f\"ur Festk\"orperforschung, Forschungszentrum \\
J\"ulich, D-52425, Germany\\
$^2$Samara State Technical University, 443010 Samara,\\
Russia}
\title{Radiative heat transfer between nanostructures}
\maketitle

\begin{abstract}
We  simplify the formalism of Polder and Van Hove  [Phys.Rev.B {\bf 4}, 3303(1971)],
which was developed to calculate the heat transfer between macroscopic and nanoscale 
bodies of arbitrary shape, dispersive and adsorptive dielectric properties. 
In the non-retarded limit, at small distances between the bodies, the  problem 
is reduced to the solution of an electrostatic problem.   We apply the 
 formalism to the study of the  heat transfer between: (a) two parallel semi-infinite bodies, 
(b) a semi-infinite body and  a spherical body, and (c) that  two spherical bodies.
 We consider the dependence of the heat transfer on the
temperature $T$, the shape and the separation $d$. We determine 
when retardation effects become important.
\end{abstract}
\section{ Introduction}
It is well known that for bodies separated by $d >>d_W\sim c\hbar/k_BT$
the radiative heat transfer between them is described by the Stefan-
Bolzman law:
\begin{equation}
J=\frac{\pi^2k_B^4}{60\hbar^3c^2}\left(T_1^4-T_2^4\right),
\end{equation}
where $T_1$ and $T_2$ are the temperatures of solid $1$ and $2$, respectively.
 In this limiting case the heat transfer  is connected with traveling 
electromagnetic waves radiated by the bodies, and does not depend on the separation 
$d$.
For $d< d_W$ the heat transfer increases by many order of magnitude, 
 which can be  explained by the existence
of evanescent electromagnetic field that decay exponentially into the vacuum.
At present time there is an increasing number of investigations of heat transfer
due to evanescent waves in connection with scanning tunneling microscopy 
and scanning thermal microscopy (STM) under ultrahigh vacuum conditions 
\cite{Van Hove,Pendry,Majumdar}. STM can be used for local heating
of the surface, resulting in local desorption or decomposition of molecular species,
and this offer further possibilities for the STM to control local chemistry on a surface.

A general formalism for evaluating the heat transfer between macroscopic bodies was
proposed some years ago by Polder and Van Hove \cite{Van Hove}. Their theory is
based on  the general theory of the fluctuating electromagnetic field developed by Rytov
 \cite{Rytov} and applied by Lifshitz \cite{Lifshitz} for studying the conservative part,
and by Volokitin and Persson \cite{Volokitin} for studying dissipative part of the
van der Waals interaction. In the present paper we  significantly simplify the formalism 
of Polder and Van Hove. We reduce the problem 
to the finding of the electromagnetic field created by a point dipole outside the     bodies. 
The formalism of Polder and Van Hove requires the determination of 
electromagnetic field for all space and for all position of a point dipole, and  
 requires
 the  integration 
 of the product of the component of the electric and magnetic 
field over the volumes of two bodies. Our formalism requires  only the evaluation
of a  surface integral over 
one of the bodies  and is simplified further in the non-retarded limit (small
 distances between bodies), where the calculation of the heat transfer is reduced to the 
problem of finding  the electrostatic potential due to a point charge. We apply the formalism 
to the calculation of the heat transfer between: (a) two semi-infinite bodies, (b) 
 a semi-infinite body and a spherical particle, and (c) two spherical particles. 
Problem (a)  was 
considered by Polder and Van Hove,
but our derivation  is much simpler. 
The problem (b) was recently studied by Pendry \cite{Pendry}. However he only 
considered large separation and neglected
 retardation effects. However for large separations retardation effects
becomes very  important.

\section{Formalism}

 Following Polder and Van  Hove 
\cite{Van Hove}, to calculate the fluctuating electromagnetic field we  use the general
theory of Rytov (see Ref. \cite{Rytov}). This
method is based on the introduction of a  fluctuating current density in  the Maxwell
equations (just as, for example, the introduction a ``random'' force in the
theory of Brownian motion of a particle). For a monochromatic field (time
factor $\exp (-\mathrm{i}\omega t)$) in a dielectric, nonmagnetic medium,
these equations are: 
\begin{eqnarray}
\mathbf{\nabla \times E} &=&i\frac \omega c\mathbf{B}  \label{zero} \\
\mathbf{\nabla }\times \mathbf{H} &=&- i\frac \omega c\mathbf{D+}%
\frac{4\pi }c\mathbf{j}^f  \label{one}
\end{eqnarray}
where, following to Rytov, we have introduced
 a fluctuating current density $\mathbf{j}^f$ associated with thermal and
quantum fluctuations.
 $\mathbf{E}$, $\mathbf{D,H}$ and$\mathbf{\,}$ $%
\mathbf{B}$ are the electric and the electric displacement field, the magnetic and the
magnetic induction fields, respectively. For nonmagnetic medium $\mathbf{B=H}
$ and $\mathbf{D}=\varepsilon \mathbf{E}$, where $\varepsilon $ is the
dielectric constant of the surrounded media. Accordingly to Rytov the average value
of the product of components of $\mathbf{j}^f$ for local optic case is given by formula
\begin{eqnarray}
\left\langle j_i^f(\mathbf{r},\omega)j_k^{f*}(\mathbf{r}^{\prime},\omega^{\prime})\right\rangle=
\left\langle j_i^f(\mathbf{r})j_k^{f*}(\mathbf{r}^{\prime})\right\rangle_\omega\delta(\omega-
\omega^{\prime}) \nonumber \\
\left\langle j_i^f(\mathbf{r})j_k^{f*}(\mathbf{r}^{\prime})\right\rangle_\omega=
A(T,\omega)\omega^2
\mathrm{Im}\varepsilon(\omega)\delta(\mathbf{r}-\mathbf{r}^{\prime})
\delta_{ik}, \label{three}  \\
A(T,\omega)=
\frac{\hbar}
{(2\pi)^2}\left(\frac 12+n(\omega)\right) \\
n(\omega) = \frac 1{e^{\hbar \omega /k_BT}-1}
\end{eqnarray}
and for nonlocal local optic
\begin{equation}
\left\langle j_i^f(\mathbf{r})j_k^{f*}(\mathbf{r}^{\prime})\right\rangle_\omega = A(T, \omega) \omega^2
\mathrm{Im}\varepsilon_{ik}(\mathbf{r},\mathbf{r}^{\prime},
\omega)
, \label{four}
\end{equation}
where $\varepsilon_{ik}(\mathbf{r},\mathbf{r}^{\prime},
\omega)$ is a non-local dielectric constant.
For simplicity, in the derivation we will assume  local optics. However, the same derivation is valid also 
for the non-local optics case, and the final results is the same in the sense that in  both
cases the problem of the heat transfer between two bodies is reduced to the  problem of finding the electromagnetic 
field outside the bodies. Compared to  Polder
and Van Hove, our treatment includes  non-local effects,
such as the anomalous skin effect. 

In order to calculate the radiative energy transfer between the bodies we  need the ensemble average
of the Poyting vector
\begin{equation}
\left\langle\mathbf{S}(\mathbf{r})\right\rangle_\omega=(c/8\pi)\left\langle\mathbf{E}(\mathbf{r})\times\mathbf{B}^*(\mathbf{r})\right\rangle_\omega + c.c.,  \label{five}
\end{equation}
at suitable point $\mathbf{r}$. From the Maxwell equations it follows that the  electric field, produced by random current density $\mathbf{j^f}$, is given by
\begin{equation}
E_i(\mathbf{r})=\int \mathrm{d} \mathbf{r}^{\prime} D_{ik}(\mathbf{r},\mathbf{r}^{\prime},\omega) j_k^f(\mathbf{r}^
{\prime}) \label{five}
\end{equation}
where function $D_{ik}(\mathbf{r},\mathbf{r}^{\prime},\omega)$ obeys the equations
\begin{equation}
\left[\nabla_i \nabla_k - \delta_{ik}( \nabla^2 - (\omega/c)^2)\varepsilon(\mathbf{r})\right]D_{kj}(\mathbf{r},
\mathbf{r}^{\prime},\omega) =(4\pi\omega\mathrm{i}
/c^2)\delta_{ij}\delta(\mathbf{r}-\mathbf{r}^{\prime})   \label{six}  
\end{equation}
\begin{equation}
\left[\nabla_j^{\prime} \nabla_k^{\prime} - \delta_{jk}( \nabla^{\prime 2} - (\omega/c)^2)\varepsilon(\mathbf{r}^
{\prime})\right]D_{ik}(\mathbf{r},
\mathbf{r}^{\prime},\omega) =(4\pi\omega\mathrm{i}
/c^2)\delta_{ij}\delta(\mathbf{r}-\mathbf{r}^{\prime})   \label{seven}  
\end{equation}

The functions $D_{ik}(\mathbf{r},\mathbf{r}^{\prime},\omega)$ have the following symmetry properties
\cite{Abrikosov}
\begin{equation}
D_{ik}(\mathbf{r},\mathbf{r}^{\prime},\omega)=D_{ki}(\mathbf{r}^{\prime},\mathbf{r},\omega)  \label{eight}
\end{equation}
The Poyting vector can be expressed trough the average products of the components of electric field.
Using equations (\ref{six},\ref{seven}) we  get
\begin{eqnarray}
\left\langle E_i(\mathbf{r}) E_j(\mathbf{r}^{{\prime}} \right\rangle_{\omega}=
 \int \mathrm{d} \mathbf{r}^{{\prime}{\prime}} A(\mathbf{r}^{{\prime}{\prime}}) \omega^2 \mathrm{Im} \varepsilon
(\mathbf{r}^{{\prime}{\prime}}) D_{ik}(\mathbf{r}, \mathbf{r}^{{\prime}{\prime}}) D_{jk}^*(\mathbf{r}^{{\prime}},
\mathbf{r}^{{\prime}{\prime}})=  \nonumber \\ 
=\frac{c^2}{2 \mathrm{i}} \int \mathrm{d} \mathbf{r}^{{\prime}{\prime}} A(\mathbf{r}^{{\prime}{\prime}})  \Bigg \{ \left[ \left( \nabla_k^{\prime
\prime} \nabla^{{\prime}{\prime}}_l-\delta_{lk} \nabla^{{\prime}{\prime}2} \right) D_{il}(\mathbf{r},
\mathbf{r}^{{\prime}{\prime}})-\frac{4\pi \omega \mathrm{i}}{c^2}\delta_{ik} \delta (\mathbf{r}-\mathbf{r}
^{\prime
\prime}) \right]  D_{jk}^{*}(\mathbf{r}^{\prime}, \mathbf{r}^{\prime \prime})   \nonumber \\ 
- \left[ \left( \nabla_k^{\prime
\prime} \nabla_l^{\prime\prime}-\delta_{lk} \nabla^{\prime\prime 2} \right) D_{jl}^*
(\mathbf{r},
\mathbf{r}^{{\prime}{\prime}})+\frac{4 \pi \omega \mathrm{i}}{c^2}\delta_{jk} \delta (\mathbf{r}
-\mathbf{r}^{\prime
\prime} \right] D_{ik}(\mathbf{r}, \mathbf{r}^{\prime \prime})\Bigg \}  \nonumber \\ 
= \frac{c^2}{2\mathrm{i}} \int \mathrm{d}\mathbf{r}^{{\prime}{\prime}}A(\mathrm{r}^{{\prime}{\prime}})\nabla^{{\prime}{\prime}}
_l \Bigg \{ 
D_{jk}^{*}(\mathbf{r}^{{\prime}},\mathbf{r}^{{\prime}{\prime}}) \left( \nabla_k^{{\prime}{\prime}} D_{il}(\mathbf{r}, \mathbf{r}^{{\prime}{\prime}})-\nabla_l^{{\prime}{\prime}} D_{ik}(\mathbf{r}, \mathbf{r}^{{\prime}{\prime}}
) \right) \nonumber \\ 
-{}D_{ik}(\mathbf{r},\mathbf{r}^{{\prime}{\prime}}) \left( \nabla_k^{{\prime}{\prime}} D_{jl}^{*}(\mathbf{r}
^{{\prime}} , \mathbf{r}^{{\prime}{\prime}})-\nabla_l^{{\prime}{\prime}} D_{jk}^{*}(\mathbf{r}^{{\prime}}, 
\mathbf{r}^{{\prime}{\prime}}
) \right)  \Bigg \} -4\omega \pi A(T_2) \mathrm{Re}D_{ij}(\mathbf{r}, \mathbf{r}^{{\prime}}) \nonumber \\ 
= \frac{(A(T_1)-A(T_2))c^2}{2\mathrm{i}} \int \mathrm{d} S_{1l}^{{\prime}{\prime}} 
\Bigg\{ D_{ik}(\mathbf{r}, \mathbf{r}^{{\prime}{\prime}}) \left(
 \nabla_l^{{\prime}{\prime}} D_{jk}^{*}(\mathbf{r}^{{\prime}},
\mathbf{r}^{{\prime}{\prime}}) -  
\nabla_k^{{\prime}{\prime}} D_{jl}^{*}(\mathbf{r}^{{\prime}},
\mathbf{r}^{{\prime}{\prime}}) \right) \nonumber \\ 
-{}
D_{jk}^{*}(\mathbf{r}^{{\prime}}, \mathbf{r}^{{\prime}{\prime}}) \left( \nabla_l^{{\prime}{\prime}} D_{ik}
(\mathbf{r}, \mathbf{r}^{{\prime}{\prime}})-
\nabla_k^{{\prime}{\prime}} D_{ji}(\mathbf{r},
\mathbf{r}^{\prime \prime}) \right) \Bigg\} 
-{}4\pi \omega A(T_2) \mathrm{Re}D_{ij}(
\mathbf{r}, \mathbf{r}^{{\prime}}),   \label{nine}
\end{eqnarray}
where we transformed the  volume integral over  bodies $1$ and $2$ to a surface integral over
 body $1$. Assume  that  the two points $\mathbf{r}$ and $\mathbf{r}^{\prime}$ lie 
outside 
the bodies. Using that for $\mathbf{r} \not= \mathbf{r}^{\prime}$
\begin{equation}
 D_{ik}(\mathbf{r}, \mathbf{r}^{{\prime}{\prime}})\nabla_k^{{\prime}{\prime}} D_{jl}^{*}(\mathbf{r}^{{\prime}},
\mathbf{r}^{{\prime}{\prime}}) =  \nabla_k^{{\prime}{\prime}}\Big( D_{ik}(\mathbf{r}, \mathbf{r}^{{\prime}{\prime}}) D_{jl}^{*}(\mathbf{r}^{{\prime
}},
\mathbf{r}^{{\prime}{\prime}})\Big), 
\end{equation}
and performing surface integral in  (\ref{nine}) gives 
\[
\left\langle E_i(\mathbf{r}) E_j(\mathbf{r}^{{\prime}} \right\rangle_{\omega} = \frac{(A(T_1)-A(T_2))c^2}{2\mathrm{i}} \int \mathrm{d}\mathbf{S}_{1}^{{\prime}{\prime}}
\cdot\Big\{ D_{ik}(\mathbf{r}, \mathbf{r}^{{\prime}{\prime}})
 \mathbf{\nabla}^{{\prime}{\prime}} D_{jk}^{*}(\mathbf{r}^{{\prime}},
\mathbf{r}^{{\prime}{\prime}}) 
\]
\begin{equation}
-{}
D_{jk}^{*}(\mathbf{r}^{{\prime}}, \mathbf{r}^{{\prime}{\prime}})  \mathbf{\nabla}^{{\prime}{\prime}} D_{ik}
(\mathbf{r}, \mathbf{r}^{{\prime}{\prime}})
 \Big\}
-{}4\pi \omega A(T_2) \mathrm{Re}D_{ij}(
\mathbf{r}, \mathbf{r}^{{\prime}}) \label{ten}
\end{equation}

Using the Maxwell equation (\ref{zero}) we can write  the Poyting vector as 
\begin{equation}
\left\langle \mathbf{S}\right\rangle_{\omega} = \frac{\mathrm{i} c^2}{8\pi \omega}\left\{ 
\mathbf{\nabla}^{\prime} 
\left\langle \mathbf{E}(\mathbf{r})\cdot
\mathbf{E}^{*}(\mathbf{r}^{\prime})\right\rangle - \left\langle \left( 
\mathbf{E}(\mathbf{r})\cdot \mathbf{\nabla}^{\prime}\right)
\mathbf{E}^{*}(\mathbf{r}^{\prime})  \right\rangle - c.c. \right\}_{\mathbf{r} = \mathbf{r}^{\prime}} 
\label{eleven}
\end{equation}

In the non-retarded limit the formalism can be simplified. In this case the
electric field can be written as the gradient of an electrostatic potential,  
 $\mathbf{E}(\mathbf{r})= - \mathbf{\nabla} \phi (\mathbf{r})$. Thus the total Poyting
vector becomes 
\begin{eqnarray*}
\left(S_{\mathit{total}}\right)_{\omega} = \frac c{8 \pi} \int \mathrm{d} \mathbf{S} 
\cdot \left\{ \left\langle
 \left[ \mathbf{E} \times  \mathbf{B}^{*} \right] \right\rangle_{\omega} + c.c. \right\} \\
=  \frac {c}{8 \pi} \int \mathrm{d} \mathbf{S} \cdot \left\{ \left\langle
- \left[
\mathbf{\nabla} \phi \times  \mathbf{B}^{*} \right] \right\rangle_{\omega} + c.c. \right\}
\\
=  \frac c{8 \pi} \int \mathrm{d} \mathbf{S} \cdot \left\{ \left\langle
 \left[
 \phi\mathbf{\nabla} \times  \mathbf{B}^{*} \right] \right\rangle_{\omega} + c.c. \right\}
\end{eqnarray*}
\begin{equation}
=  \frac {\mathrm{i} \omega}{8 \pi} \int \mathrm{d} \mathbf{S} \cdot \mathbf{\nabla}^{\prime} \left( \left\langle
 \phi (\mathbf{r})\phi^{*} (\mathbf{r}^{\prime})  \right\rangle_{\omega} - c.c. \right)_{\mathbf{r}=
\mathbf{r}^{\prime}} \label{twelve}
\end{equation}
In the same approximation we can write
\[
D_{ik}(\mathbf{r}, \mathbf{r}^{\prime})= -\frac {\mathrm{i}}{\omega} \nabla_i \nabla_k^{\prime}
D(\mathbf{r}, \mathbf{r}^{\prime}), 
\]
where the function $D(\mathbf{r}, \mathbf{r}^{\prime})$ obeys the Poisson's equation
\begin{equation}
\Delta D(\mathbf{r}, \mathbf{r}^{\prime})= -4\pi \delta (\mathbf{r}- \mathbf{r}^{\prime}) \label{fourteen} 
\end{equation}
Using the identities
\begin{eqnarray}
D_{ik}(\mathbf{r}, \mathbf{r}^{\prime \prime}) \left( \nabla_l^{\prime \prime} D_{jk}^{*}
(\mathbf{r}^{\prime}, \mathbf{r}^{\prime \prime}) - \nabla_k^{\prime \prime} D_{jl}^{*}
(\mathbf{r}^{\prime}, \mathbf{r}^{\prime \prime})\right) \nonumber  \\
=- \frac{\mathbf{i}}{\omega}\nabla_i \Bigg\{ \nabla_k^{\prime \prime} \left[ D(\mathbf{r}, \mathbf{r}^
{\prime \prime}) \left( \nabla_l^{\prime \prime} D_{jk}^{*}
(\mathbf{r}^{\prime}, \mathbf{r}^{\prime \prime}) - \nabla_k^{\prime \prime} D_{jl}^{*}
(\mathbf{r}^{\prime}, \mathbf{r}^{\prime \prime})\right) \right] - \nonumber \\
-  D(\mathbf{r}, \mathbf{r}^{\prime \prime})
\left( \nabla_l^{\prime \prime} \nabla_k^{\prime \prime} D_{jk}^{*}
(\mathbf{r}^{\prime}, \mathbf{r}^{\prime \prime}) - \nabla_k^{\prime \prime 2} D_{jl}^{*}
(\mathbf{r}^{\prime}, \mathbf{r}^{\prime \prime})\right) \Bigg\} \nonumber \\
= - \frac{\mathrm{i}}{\omega}\nabla_i \nabla_k^{\prime \prime} \left[ D(\mathbf{r}, \mathbf{r}^
{\prime \prime}) \left( \nabla_l^{\prime \prime} D_{jk}^{*}
(\mathbf{r}^{\prime}, \mathbf{r}^{\prime \prime})  
 - \nabla_k^{\prime \prime} D_{jl}^{*}
(\mathbf{r}^{\prime}, \mathbf{r}^{\prime \prime})\right) \right]  \nonumber  \\  
 - \frac 1{c^2}\nabla_i \nabla_j^{\prime}
 D(\mathbf{r}, \mathbf{r}^
{\prime \prime})\nabla_l^{\prime \prime}D^{*}(\mathbf{r}^{\prime}, \mathbf{r}^
{\prime \prime}), \label{thirteen}
\end{eqnarray}
formula (\ref{nine}) gives 
\begin{eqnarray}
\left\langle E_i(\mathbf{r}) E_j(\mathbf{r}^{{\prime}} \right\rangle_{\omega} = \nabla_i 
\nabla_j^{\prime} \left\langle \phi (\mathbf{r})  \phi ^{*}(\mathbf{r}^{\prime}) \right\rangle_{\omega}  \\
 \left \langle \phi (\mathbf{r})  \phi^{*} (\mathbf{r}^{\prime}) \right\rangle_{\omega} = \frac{A(T_1)
-A(T_2)}{2\mathrm{i}} \int \mathrm{d} S_{1l}^{\prime \prime} 
\Bigg\{ D^{*}(\mathbf{r}^{\prime}, \mathbf{r}^
{\prime \prime}) \nabla_l^{\prime \prime} D(\mathbf{r}, \mathbf{r}^
{\prime \prime})   \nonumber  \\
 - D(\mathbf{r}, \mathbf{r}^
{\prime \prime}) \nabla_l^{\prime \prime} D^{*}(\mathbf{r}^{\prime}, \mathbf{r}^
{\prime \prime})\Bigg\} 
- 4\pi A(T_2)\mathrm{Im}D(\mathbf{r}, \mathbf{r}^
{\prime }). \label{aone} 
\end{eqnarray}

\section{Heat transfer between two flat surfaces}

In this section we apply the general formalism,  developed in the previous
section, to the problem of the heat transfer between two flat surfaces. This problem
was considered some years ago by Polder and Van Hove \cite{Van Hove}, and 
the non-retarded limit was  recently studied
by Pendry \cite{Pendry}. The present
derivation is more compact and demonstrate the elegance of our general
approach. We  suppose that the half-space $z<0$ is filled by a medium (temperature $T_1$) 
with
reflection factors $R_{1p}(\mathbf{q},\omega)$ and $R_{1s}(\mathbf{q},\omega)$
for $s$- and $p$- polarized electromagnetic fields, respectively, 
 and the half-space $z>d$ is filled by a medium (temperature $T_2$) with 
reflection factors $R_{2p}(\mathbf{q},\omega)$ and  $R_{2s}\mathbf{q},\omega)$, and 
the
region between the solids, $0<z<d$, is assumed to be vacuum. Let $\mathbf{q}$ be  
the component of wave vector $\mathbf{k}=(\mathbf{q},p)$
parallel to the surfaces and
\begin{equation}
p=\sqrt{\left( \frac \omega c\right) ^2 -q^2}            
\end{equation}
Since the problem is translationaly invariant in the $\mathbf{x}=(x,y)$ plane, the Green's
function $D_{ij}(\mathbf{r}, \mathbf{r}^{{\prime}})$ can be represented by the
Fourier integral:
\begin{equation}
D_{ij}(\mathbf{r}, \mathbf{r}^{\prime}) = \int \frac{\mathrm{d}^2q}{(2\pi)^2}
e^{\mathrm{i}\mathbf{q}\cdot(\mathbf{r} - \mathbf{r}^{\prime})}D_{ij}(z, z^{\prime},
\mathbf{q}, \omega),
\end{equation}
Taking the $x$- axis along the vector $\mathbf{q}$,  equations (\ref{six}) and (\ref{seven})
become 
\begin{eqnarray}
\left(p^2 + \frac{\partial^2}{\partial z^2}\right)D_{yy}(z, z^{\prime})
 = - \frac{4\pi \mathrm{i} \omega}
{c^2} \delta(z-z^{\prime}) \label{fifteen} \\
\left(p^2 + \frac{\partial^2}{\partial z^2}\right)D_{xx}(z, z^{\prime}) - \mathrm{i}q\frac{\partial}
{\partial z}D_{zx}(z, z^{\prime}) = - \frac{4\pi  \mathrm{i}  \omega}
{c^2} \delta(z-z^{\prime}) \label{sixteen} \\
p^2D_{zx}(z, z^{\prime})-  \mathrm{i}q\frac{\partial}
{\partial z}D_{xx}(z, z^{\prime}) = 0 \label{seventeen} \\
p^2D_{zz}(z, z^{\prime}) - \mathrm{i}q\frac{\partial}
{\partial z}D_{xz}(z, z^{\prime} = -\frac{4\pi  \mathrm{i}  \omega}
{c^2} \delta(z-z^{\prime}) \label{eighteen} \\
p^2D_{xz}(z, z^{\prime}) + \mathrm{i}q\frac{\partial}
{\partial z^{\prime}}D_{xx}(z, z^{\prime}) = 0  \label{nineteen} 
\end{eqnarray}
Since the equations for  $D_{xy}$  and $D_{zy}$ are  homogeneous, these  components of the Green function
must vanish. 
Thus solving the system of equations (\ref{fifteen}) - (\ref{nineteen}) is equivalent to solving altogether two equations: 
equation (\ref{fifteen}) for $D_{yy}$,
and the equation for $D_{xx}$ which follows from equations (\ref{sixteen}) and (\ref{seventeen}):
\begin{equation}
\left(p^2 + \frac{\partial^2}{\partial z^2}\right)D_{xx}(z, z^{\prime})
 = - \frac{4\pi \mathrm{i} p^2}
{\omega} \delta(z-z^{\prime}) \label{twenty}. \\
\end{equation}
$D_{xz}$, $D_{zx}$ and $D_{zz}$ for $z\not=z^{\prime}$ can be obtained as
\begin{equation}
D_{xz}=-\frac{\mathrm{i}q}{p^2}\frac{\partial}{\partial z^{\prime}}D_{xx}  ;\; D_{xz}=-\frac{\mathrm{i}q}
{p^2}
\frac{\partial}{\partial z^{\prime}}D_{xx}  ; \;  
 D_{zz} = \frac {q^2}{p^4}\frac {\partial^2}{\partial z \partial
z^{\prime}} D_{xx}
\end{equation}
In the vacuum gap, $0<z<d$, the solution of  (\ref{fifteen}) has the form
\begin{equation}
D_{yy}(z, z^{\prime}) = - \frac {2 \pi \omega}{pc^2}e^{\mathrm{i}p\mid z-z^{\prime}\mid} +v_ye^
{\mathrm{i}pz} + w_ye^
{-\mathrm{i}pz}  \label{twentyone}
\end{equation}
At the boundaries  $z=0$ and $z=d$, the amplitude of the reflected wave is equal to the
amplitude  of the incident wave times to corresponding reflection factor. The Green's  function $D_{yy}$ is 
associated with the $s$-wave, which satisfies
the boundary conditions 
\begin{eqnarray}
v_y = R_{1s} \left(w_y- \frac {2\pi \omega}{pc^2}e^{\mathrm{i}pz^{\prime}}\right)
\qquad\mbox{for}\qquad
z=0  \label{sixty}\\
w_y = R_{2s}e^{2\mathrm{i}pd} \left(v_y- \frac {2\pi \omega}{pc^2}e^{-\mathrm{i}pz^{\prime}}\right)
\qquad\mbox{for}\qquad z=d \label{sixtyone}
\end{eqnarray}
Using (\ref{twentyone})-(\ref{sixtyone})
\begin{eqnarray}
D_{yy}(z, z^{\prime})=-\frac{2\pi \omega}{pc^2}\Bigg\{e^{\mathrm{i}p\mid z-z^{\prime}\mid} 
+\frac{R_{1s}R_{2s}e^{2\mathrm{i}pd}\left(e^{\mathrm{i}p(z-z^{\prime})}+ 
e^{-\mathrm{i}p(z-z^{\prime})}\right)}{\Delta_s}  \nonumber  \\
+\frac {R_{1s}e^{\mathrm{i}p(z+z^{\prime})} + R_{2s}e^{2\mathrm{i}pd} e^{-\mathrm{i}p(z+z^{\prime})}}
{\Delta_s}
\Bigg\}  \label{twentytwo}  \\
\Delta_s = 1 - e^{2\mathrm{i}pd}R_{2s}R_{1s}
\end{eqnarray}
Equation (\ref{twenty}) for $D_{xx}$ is similar to equation (\ref{fifteen}) for  $D_{yy}$,
and  taking into account that $D_{xx}$ is associated with $p$- wave it can be obtained directly from  (\ref{twentytwo})  by replacing
$R_s\rightarrow R_p$:
\begin{equation}
 D_{xx} = \left(\frac{pc}{\omega}\right)^2D_{yy}\left[R_s \rightarrow R_p \right] \label{twentyeight}
\end{equation}
We  note that in our approach the calculation of the reflection factors for $s$- and $p$- waves
is considered as separated problem, which , if necessary, can be solved by taking into account non-local effects. 
 For the local optic case the reflection factors are determined by the well known Fresnel formulas
\begin{equation}
R_{ip}=\frac{\varepsilon _ip- s_i}{\varepsilon _ip+ s_i}
,\,\,\;\;\;\;\;R_{is}=\frac{p-s_i}{p+s_i},  \label{twentyfive}
\end{equation}
where $\varepsilon _i\,$ is the complex dielectric constant for body 
$i$, 
\begin{equation}
s_i=\sqrt{\frac{\omega ^2}{c^2}\varepsilon _i-q^2}.  \label{twentyfour}
\end{equation}
For homogeneous (in $xy$-plane) bodies the second term in  (\ref{eleven})
for $z$- component for  Poyting vector can be written in the form
\[
\nabla_k^{\prime}\left\langle E_k(\mathbf{r})
E_z^{*}(\mathbf{r}^{\prime})\right\rangle_{\mathbf{r}=\mathbf{r}^{\prime}} = \nabla_k\left\langle 
E_k(\mathbf{r})
E_z^{*}(\mathbf{r})\right\rangle = \nabla_z\left\langle 
E_z(\mathbf{r})
E_z^{*}(\mathbf{r})\right\rangle
\]
The last quantity in this expression is purely real, and does not give any contribution 
to $z$ - component of the Poyting vector.
Thus, from equations (\ref{ten}) and (\ref{eleven}) we obtain
\begin{eqnarray} 
\left\langle S_z\right\rangle_{\omega} = \frac{c^4\left[A(T_1)-A(T_2)\right]}{16\pi \omega}
\int \frac{\mathrm{d}^2\mathbf{q}}{(2\pi)^2}\Bigg\{  \Bigg(D_{yy}\frac{\partial^2}{\partial z
\partial z^{\prime}}D^{*}_{yy}(z, z^{\prime}) - \nonumber  \\ 
\frac{\partial}{\partial z}D_{yy} 
\frac{\partial}{\partial z^{\prime}}D_{yy}(z, z^{\prime})\Bigg) +  \left(\frac{\omega}
{pc}\right)^4\left( \left[ y \rightarrow x \right] \right) + c.c. \Bigg\}_{z\rightarrow z^{\prime}+0}
\label{twentyseven}
\end{eqnarray}
Using (\ref{twentytwo}) and (\ref{twentyeight}) in (\ref{twentyseven}) we finally obtain
\begin{eqnarray}
S_z &=&\frac \hbar {8\pi ^3}\int_0^\infty \mathrm{d}\omega \omega
\int_{q<\omega /c }\mathrm{d}^2q  \nonumber \\
&&\times\left[
\frac{(1-\mid R_{1p}(\omega )\mid ^2)(1-\mid R_{2p}(\omega)\mid ^2)%
(n_{1}(\omega )-n_{2}(\omega))}{\mid 1-%
\mathrm{e}^{2\mathrm{i}pd}R_{1p}(\omega )R_{2p}(\omega )\mid ^2}\right]
\nonumber \\
&&  +\frac \hbar {2\pi
^3}\int_0^\infty \mathrm{d}\omega \omega\int_{q>\omega /c}%
\mathrm{d}^2q\mathrm{e}^{-2\mid p\mid d}  \nonumber \\
&&\times  \frac{\mathrm{Im}R_{1p}(\omega )\mathrm{Im}%
R_{2p}(\omega )}{\mid 1-\mathrm{e}^{-2\mid p\mid d}R_{1p}(\omega
)R_{2p}(\omega)\mid ^2}(n_{1}(\omega)-n_{2}(\omega ))
\nonumber  \label{fourtyfour} \\
&&\left. +\left[ p\rightarrow s\right]  \label{seventyone} \right.
\end{eqnarray}
where
\begin{equation}
n_1(\omega) = \left (e^{\hbar \omega /k_BT_1}-1\right )^{-1}
\end{equation}
is the Bose-Einstein factor of solid \textbf{1} and similar for $n_2$.

Fig. 1a shows the heat transfer between two semi-infinite silver bodies
separated
by the distance $d$ and at the temperatures
$T_1 = 273 {\rm K}$ and $T_2 = 0 {\rm K}$, respectively. The $s$- and $p$-wave contributions
are shown separately,
and the $p$-wave contribution has been calculated using non-local optics (the
dashed
line shows the result using local optics). It is remarkable how important the
$s$-contribution is even for short distances.
The detailed distance dependence of $S_z$ has been studied
by Polder and Van Hove within the local optics approximation, and will not
be repeated here. The nonlocal optics contribution to $(S_z)_p$,
which is important only for $d < l$ (where $l$ is the electron mean free path
in the bulk),
is easy to calculate for free electron like metals.
The non-local surface contribution to ${\rm Im} R_p$ is given by \cite{Persson
and Zhang}
$$\left ( {\rm Im} R_p \right )_{\rm surf} = 2 \xi {\omega\over \omega_p}
{q\over k_F},$$
where $\xi(q)$ depends on the electron density parameter $r_s$ but
typically $\xi(0)\sim1$. 
Using this expression for ${\rm Im} R_p$ in (\ref{seventyone}) gives the
(surface)
contribution:
$$S_{\rm surf} \approx { \xi^2 k_B^4 \over \omega_p^2 k_F^2 d^4 \hbar^3 }
\left(T_1^4-T_2^4\right).$$
Note from Fig. 1a that the local optics
contribution to
$S_{pz}$ depends nearly linearly on $1/d$ in the studied distance interval,
and that this
contribution is much smaller than the $s$-wave contribution. Both these
observations
differ from Ref. \cite{Pendry}, where it is stated that the $s$ contribution
can be
neglected for
small distances and that the $p$-wave contribution (within local optics) is
proportional to
$1/d^2$ for small distances. However, for  very high-resistivity
materials, the $p$-wave contribution becomes much more important,
and a crossover to a $1/d^2$-dependence of $S_{pz}$ is observed at  short
separations
$d$. This is illustrated in Fig. 1b and 1c, which have been calculated with the
same
parameters as in Fig. 1a, except that the electron mean free path has been
reduced from
$l= 560 \ {\rm \AA}$ (the electron mean free path for silver at room
temperature)
to $20 \ {\rm \AA}$ (roughly the electron mean free path in lead at room
temperature)
(Fig. 1b) and $3.4 \ {\rm \AA}$
(of order the lattice constant, representing the minimal possible mean free
path)
(Fig. 1c). Note that when $l$ decreases, the $p$ contribution to the heat
transfer increases while
the $s$ contribution decreases. Since the mean free path cannot be much smaller
than the lattice constant, the result in Fig. 1c
represent the largest possible $p$-wave
contribution for normal metals. However, the $p$-wave contribution may be even
larger for
other materials, e.g., semimetals, with lower carrier concentration than in
normal metals.
This fact has already been pointed out by Pendry: the p-wave contribution for
short distances
is expected to be maximal when
the function
$${\rm Im} R_p \approx {\rm Im} {\epsilon -1 \over \epsilon +1}
={\rm Im} \left [ 1-2{\omega \over \omega_p}\left ({\omega \over
\omega_p}+{i\over \omega_p \tau}
\right )\right ]^{-1}$$
is maximal with respect to variations in $1/\tau$. This gives:
$$\omega_p \tau = {2k_BT \over \hbar \omega_p}$$
where we have used that typical frequencies $\omega \sim k_B T /\hbar$. Since
the
DC resistivity $\rho = 4\pi/(\omega_p^2\tau)$ we get (at room temperature)
$\rho \approx 2 \pi \hbar /k_BT
\approx 0.14 \ \Omega {\rm cm}$.
To illustrate this case,
Fig. 2 shows the thermal flux as a function of the separation $d$ between the
surfaces
when $(\hbar /\tau)/k_BT
=120$ and $\hbar \omega_p /k_BT = 15.5$.
One body is at zero temperature and the other at $T=273K$.

Fig. 3 shows the thermal flux as a function of the resistivity of the solids.
Again we assume that
one body is at zero temperature and the other at $T=273K$.
The solid surfaces are separated by $d = 10 \ {\rm \AA}$ and $\hbar \omega_p >>
k_BT$.
The heat flux for other separations can be obtained using scaling $\sim 1/d^2$
which
holds for high-resistivity materials.
Finally, we note that thin high-resistivity coatings can drastically increase
the
heat transfer between two solids. This is illustrated in Fig. 4 shows the heat
flux
when thin films ($\sim 10\,\mathrm{\AA}$) of high resistivity material $\rho = 0.14 \ \Omega {\rm cm}$,
are deposited
on silver.
One body is at zero temperature and the other at $T=273K$.
(a) and (b) shows the $p$ and $s$-contributions, respectively. Also shown are
the
heat flux when the two bodies are made from silver,
and from the high resistivity material.
It is interesting to note that while the $p$-contribution
to the
heat flux for the  coated surfaces is strongly influenced by
the coating, 
 the $s$-contribution is nearly unaffected.

\section{Local heating of a surface by an STM tip}
It was pointed by  Pendry \cite{Pendry} the local heating of a surface 
by STM tip can be used for local modification of a surface if the heat transfer 
is sufficiently great. To investigate the power of a hot tip to heat a surface
Pendry modeled the tip as a hot sphere of the same radius $R$ as the tip. This is a common
approximation when calculating tunneling current  and the same arguments justify its use
for calculating heat tunneling. Pendry considered the case $R<<d<<d_W \sim c\hbar/k_BT$
and the electrostatic limit. However at large distances retardation effects can be
important. In fact,  it will be shown below that the heat transfer between a sphere and
surface in the asymptotic limit (large separation) can be obtained directly
 from formula (\ref{seventyone}).

Consider distances $d<<d_W\sim c\hbar /k_BT$ (at $T=300\;K$ \ we have\ \ $d_W$
$\sim 10^5\,\mathrm{\AA }$). In this case we can neglect by the first
integral in (\ref{fourtyfour}), put $p\approx \mathrm{i}q$, and extend the
  integral  to the whole $q$ -  plane. Using these
approximations, the second integral in (\ref{seventyone}) can be written as
\begin{eqnarray}
S_z =\frac{\hbar }{\pi ^2}\int_0^\infty \mathrm{d}\omega \omega \left(n_1(\omega)-n_2(\omega)\right) 
\int_0^\infty \mathrm{d}qq\mathrm{e}
^{-2qd}   \nonumber  \\
\times \left\{ \frac{\mathrm{Im}R_{1p}(\omega )\mathrm{Im}
R_{2p}(\omega )}{\mid 1-\mathrm{e}^{-2qd}R_{1p}(\omega )R_{2p}(\omega )\mid
^2}+[p\rightarrow s]\right\}. 
  \label{fourtyeight}
\end{eqnarray}
Now, assume that the medium ${\bf 2}$ is sufficiently rarefied and
consist of  particles with the radius $R<<d$, and the polarizability   $\alpha(\omega)$. Then
 $\varepsilon_2 - 1 \rightarrow 4\pi \alpha_2 n<<1$, where
$n$ is the number of particles per unit volume. Thus, when $n\rightarrow 0$
it is enough to include only the first-
non-vanishing terms in the expansion of the integrand of Eq.(\ref{fourtyeight})
in powers of  $\varepsilon_2 - 1$. The heat transfer between one particle
and a surface can be obtained as the ratio between the change of
heat transfer after displacement of body ${\bf 2}$ by small distance $\mathrm{d}z$,
and the number of particle in a slab
with thickness $\mathrm{d}z$: 

\begin{eqnarray}
S_z =2\frac{\hbar }{\pi }\int_0^\infty \mathrm{d}\omega \omega 
\left(n_1(\omega)-n_2(\omega)
\right)
\int_0^\infty \mathrm{d}qq^2\mathrm{e}
^{-2qd}  \nonumber \\
\times \left\{ 2\mathrm{Im}R_{1p}(\omega )\mathrm{Im}
\alpha_{2}(\omega )
+\left(\frac{\omega}{cq}\right)^2\mathrm{Im}R_{1s}(\omega )\mathrm{Im}
\alpha_{2}(\omega  \right\} 
  \label{fourtynine}
\end{eqnarray}

To treat this expression we assume that $\mid \varepsilon(\omega)\mid>>1$ holds for all
relevant frequencies. In the limit $d<\mid \varepsilon \mid^{-1/2}d_W$, where $\varepsilon$
is taken at the characteristic frequency $\sim k_BT/\hbar$,
 the reflection factor of $p$-wave becomes 
\begin{equation}
R_{1p}\approx\frac{\varepsilon_1-1}{\varepsilon_1+1}, \;\;\;\mathrm{Im}R_{1p}\approx
\frac{2\mathrm{Im}\varepsilon_1}{\mid \varepsilon_1\mid^2}. \label{dthree}
\end{equation} 
The  polarizability  of the sphere is determined by 
\begin{equation}
\alpha_2=\frac{\varepsilon_2-1}{\varepsilon_2+2}R^3, \;\;\;\mathrm{Im}\alpha_2\approx
\frac{3\mathrm{Im}\varepsilon_2}{\mid \varepsilon_2\mid^2}. \label{sixty}
\end{equation}
We  describe the sphere  the same dielectric function as
the substrate:
\begin{equation}
\varepsilon (\omega )=1-\frac{\omega _{p}^2}{\omega (\omega +\mathrm{i}/\tau )}%
,  \label{sixtythree}
\end{equation}
where $\tau $ is the Drude relaxation time and $\omega _p$ the plasma
frequency.
In this case  the  $p$-wave contribution becomes 
\begin{equation}
S_{pz} \approx \frac{2\pi^3 R^3 k^4_B }{5d^3\hbar^3}\frac 1{(\omega_p^2\tau)^2}(T_1^4-T_2^4) 
\label{fifty}
\end{equation}
This result is in an agreement with calculations of Pendry \cite{Pendry}.
To evaluate the  $s$-wave  contribution in the limit
  $d<\mid \varepsilon \mid^{-1/2}d_W$, we use the integral
\begin{eqnarray*}
\int_0^\infty \mathrm{d}q e^{-2qd}\mathrm{Im}\frac{\mathrm{i}q-s}{\mathrm{i}q+s}\approx
\int_0^\infty \mathrm{d}q\mathrm{Im}\frac{\mathrm{i}q-s}{\mathrm{i}q+s}  \\
=\mathrm{Im}\left\{\mid s_0\mid e^{\mathrm{i}\phi} \int_0^\infty \mathrm{d}t
\frac{t-\sqrt{t^2-1}}{t+\sqrt{t^2-1}}\right\}=  \\
=\frac 1{2}\mathrm{Im}\left\{\mid s_0\mid e^{\mathrm{i}\phi} \int_{-\mathrm{i}\pi/2}^\infty \mathrm{d}z
\left(e^{-z}-e^{-3z}\right)\right\}=\frac {2}{3}\mid s_0\mid \cos(\phi),
\end{eqnarray*}
where $s_0=s(q=0)=(\omega/c)\sqrt{\varepsilon}, \mid s_0\mid^{1/2}, \phi=\arg{s_0}$.
Thus,
\begin{equation}
S_{sz}\approx 2\cdot10^2k_B^{11/2}\hbar^{-9/2}R^3c^{-3}T^{11/2}(\omega_p^2\tau)^{-1/2}.
\label{fiftyone} 
\end{equation}  
From the comparison (\ref{fifty}) and (\ref{fiftyone}) the $s$-wave contribution exceeds
the $p$-wave contribution for $d>(d_W  c/\omega_p^2 \tau)^{1/2}$ 
; for typical metals at room temperature, $\hbar \omega_p/k_BT \sim 10^3$ and 
$\omega_p \tau \sim 10^2$ so that $d>10^2\,\mathrm{\AA}$. For  $d_W\mid \varepsilon \mid^{-1/2} <d<
 d_W$ we obtain
\begin{equation}
S_{pz}\approx 10k_B^{9/2}\hbar^{-7/2}d^{-2}R^3c^{-1}T^{7/2}(\omega_p^2\tau)^{3/2}.
\end{equation}
For  $
 d_W\mid \varepsilon \mid^{-1/2}<d<d_W$ we obtain the  $s$-wave contribution
\begin{equation}
S_{sz}\approx 10k_B^{9/2}\hbar^{-7/2}d^{-2}R^3c^{-1}T^{9/2}(\omega_p^2\tau)^{-3/2}.
\end{equation}

Assume now that the spherical particle is so close to the surface that we can neglect
 retardation effects. In this case the problem is reduced to  finding
of electrostatic potential created by point charge, located in vacuum. Using image
theorem \cite{Landau} the electrostatic potential can be written in the form
\begin{eqnarray}
D(\mathbf{r}, \mathbf{r}^{\prime})= \frac 1{\mid \mathbf{r}- \mathbf{r}^{\prime}
\mid}-\frac{\varepsilon_1-1}{\varepsilon_1+1}\frac 1{\mid \mathbf{r}- 
\tilde{\mathbf{r}}^{\prime}\mid} + \nonumber \\
\sum_{m=0}^\infty \sum_{n=m}^\infty C_{n}^m(\mathbf{r}^{\prime})\left(\frac {P_n^m(\cos \theta)}
{r^{n+1}} - (-1)^{n+m} \frac{\varepsilon_1-1}{\varepsilon_1+1}\frac {P_n^m(\cos \theta_i)}
{r_i^{n+1}}\right)\cos m(\phi - \phi^{\prime}) \label{none} 
\end{eqnarray}
where we have choosen the origin of the coordinate system at the center $O$ of the spherical
particle, and taken  the polar axis along the line connecting  $O$ with
the center of ``image" sphere $O^{\prime}$, and assumed that the points $\mathbf{r}$
and $\mathbf{r}^{\prime}$
have the polar coordinates $(r, \theta, \phi)$ and $(r^{\prime}, \theta^{\prime},
 \phi^{\prime})$ with respect  $O$; 
$\tilde{\mathbf{r}}^{\prime}$ is the ``image" of  $\mathbf{r}^{\prime}$
; $\mathbf{r}_i=\mathbf{r}-\mathbf{h}=(r_i, \theta_i, \phi)$ where $\mathbf{h}$ is
the vector connecting the centers $O$ and $O^{\prime}$. $h=2(R+d)$ and
$P_n^m(\cos \theta)$ is the associated Legendre function (see Fig.5).
At any interior point of the sphere the resultant potential is
\begin{equation}
D(\mathbf{r}, \mathbf{r}^{\prime})=\sum_{m=0}^\infty \sum_{n=m}^\infty A_n^m(\mathbf{r}^{\prime})
r^nP_n^m(\cos \theta)\cos m(\phi - \phi^{\prime}) \qquad (r<R)
\end{equation}
We expand the potential (\ref{none}) in the spherical harmonics around the center
$O$  using the formulas \cite{Nozawa}
\begin{eqnarray}
\frac 1{\mid \mathbf{r}- \mathbf{r}^{\prime}\mid}= 
\sum_{m=0}^\infty \sum_{n=m}^\infty
\frac {r^n}{r^{\prime n+1}}\frac{(n-m)!}{(n+m)!} \nonumber \\
\times P_n^m(\cos \theta)P_n^m(\cos \theta^{\prime})
\epsilon_m\cos m(\phi - \phi^{\prime}) \qquad (r<r^{\prime}) \label{ntwo} \\
\frac 1{\mid \mathbf{r}-
\tilde{\mathbf{r}}^{\prime}\mid} 
=\sum_{m=0}^\infty \sum_{n=m}^\infty\sum_{l=m}^\infty (-1)^{l+n}
\frac{r^n r^{\prime l}}{h^{l+n+1}} \frac {(l+n)!}{(l+m)!(n+m)!} \nonumber \\
\times P_n^m(\cos \theta)P_l^m(\cos \theta^{\prime})
\epsilon_m\cos m(\phi - \phi^{\prime}) \qquad (r,r^{\prime}<h/2) \\
(-1)^{n+m} \frac{P_n^m(\cos \theta_i)}
{ir_i^{n+1}}=\sum_{l=m}^\infty (-1)^{l+n}\frac{r^l}{h^{n+l+1}}{l+n\choose l+m}P_l^m(\cos \theta)
\qquad (r<h) \label{nthree}
\end{eqnarray}
where $\epsilon_0=1, \epsilon_m=2$ for $m\not=0$.
Using (\ref{ntwo})-(\ref{nthree}) we can rewrite  (\ref{none}) in the form
\begin{eqnarray}
D(\mathbf{r}, \mathbf{r}^{\prime})=\sum_{m=0}^\infty \sum_{n=m}^\infty\sum_{l=m}^\infty
\Bigg\{r^n\Bigg(\frac 1{r^{\prime n+1}}\frac{(n-m)!}{(n+m)!}\delta_{nl}+ \nonumber \\
(-1)^{l+n+1}\frac{\varepsilon_1-1}{\varepsilon_1+1}
\frac{ r^{\prime l}}{h^{l+n+1}} \frac {(l+n)!}{(l+m)!(n+m)!}\Bigg) 
P_n^m(\cos \theta)P_l^m(\cos \theta^{\prime}) \nonumber \\
+  C_{n}^m(\mathbf{r}^{\prime})\Bigg(\frac {\delta_{nl}}{r^{n+1}} 
+(-1)^{l+n+1}\frac{\varepsilon_1-1}{\varepsilon_1+1}
\frac{r^l}{h^{l+n+1}} {l+n\choose l+m} \Bigg)
 P_l^m(\cos \theta)
\Bigg\}\epsilon_m\cos m(\phi - \phi^{\prime}) \label{nnone}
\end{eqnarray}
Across the surface of the sphere the Green function $D(\mathbf{r}, \mathbf{r}^{\prime})$
(as the function of $\mathbf{r}$), and its normal derivatives must be continuous.
These boundary conditions lead to the equation
\begin{eqnarray}
C_n^m(\mathbf{r}^{\prime})=A_n\sum_{l=m}^\infty\Bigg\{\Bigg(\frac 1{r^{\prime n+1}} 
\frac{(n-m)!}{(n+m)!}\delta_{nl} \nonumber \\
+(-1)^{l+n+1}\frac{\varepsilon_1-1}{\varepsilon_1+1}
\frac{ r^{\prime l}}{h^{l+n+1}} \frac {(l+n)!}{(l+m)!(n+m)!}\Bigg)
P_l^m(\cos \theta^{\prime})  \nonumber \\
+\frac{\varepsilon_1-1}{\varepsilon_1+1}
\left(-\frac {1}{h}\right)^{l+n+1} {n+l\choose n +m}C_l^m(\mathbf{r}^{\prime})\Bigg\} \label{nfour} \\
A_n=\frac{(1-\varepsilon_2)n}{n\varepsilon_2+n+1}R^{2n+1}
\end{eqnarray}
Outside the sphere the function $C_n^m(\mathbf{r}^{\prime})$ satisfies to Laplace's
equation and can be expanded as
\begin{equation}
C_n^m(\mathbf{r}^{\prime}) =\sum_{l=m}^\infty\left(\frac{a_{nl}^m}{r^{\prime \,l+1}}
+b_{nl}^mr^{\prime \,l}\right)P_l^m(\cos \theta^{\prime}) \label{nfive}
\end{equation} 
Using (\ref{nfive}) in (\ref{nfour}) and taking into account that equation
(\ref{nfour}) must be satisfied at arbitrary $r^{\prime}$ we obtain equations
for coefficients $a_{nl}^m$ and $b_{nl}^m$
\begin{eqnarray}
a_{nl}^m  =A_n\sum_{l^{\prime}=m}^\infty\Bigg\{
\frac{(n-m)!}{(n+m)!}\delta_{nl^{\prime}}\delta_{ll^{\prime}}+
\frac{\varepsilon_1-1}{\varepsilon_1+1}
\left(-\frac {1}{h}\right)^{l^{\prime}+n+1} {n+l^{\prime}\choose n+m}
a_{l^{\prime}l}^m\Bigg\}  \label{nsix} \\
b_{nl}^m=A_n\frac{\varepsilon_1-1}{\varepsilon_1+1}  \sum_{l^{\prime}=m}^\infty
 \left( -\frac {1}{h}\right)^{l^{\prime}+n+1} \left( 
\frac {(l^{\prime}+n)!}{(l^{\prime}+m)!(n+m)!}\delta_{ll^{\prime}}+
{n+l^{\prime}
\choose n+m}b_{l^{\prime}l}^m \right) \label{nseven}
\end{eqnarray}
If we make  the replacement
\begin{equation}
a_{nl}^m = A_n \frac{(n-m)!}{(n+m)!}\delta_{nl} + \tilde{a}_{nl}^m,
\end{equation}
then from (\ref{nsix}) and (\ref{nseven}) we obtain
\begin{equation}
b_{nl}^m = \frac{\tilde{a}_{nl}^m}{A_l}.
\end{equation}
Let us introduce the dimensionless coefficients $x_{nl}$
\begin{eqnarray}
\tilde{a}_{nl}^m = \left( -\frac {1}{h}\right)^{l+n+1}
\frac{R_{1p}}{(n+m)!(l+m)!}A_n A_l x_{nl} = \nonumber  \\
(-\xi R)^{n+l+1}\frac{R_{1p}}{(n+m)!(l+m)!}\lambda_n \lambda_lx_{nl} ,
\end{eqnarray}
where
\[ 
R_{ip}=\frac {\varepsilon_i-1}{\varepsilon_i+1},
\]
$\xi=(R/h), \lambda_n = -A_n/R^{2n+1}$. The coefficients $x_{nl}$ obey 
the equation
\begin{equation}
x_{nl} = (n+m)! + R_{1p}\sum_{l^{\prime}=m}^\infty \xi^{2l^{\prime}+1}
\lambda_{l^{\prime}}\frac{(n+l^{\prime})!}{(l^{\prime}-m)!(l^{\prime}+m)!}
x_{l^{\prime}l} \label{bone}
\end{equation}
The solution of (\ref{bone}) can be found by iterations and has
the form
\begin{eqnarray}
x_{nl} = (n+m)! + R_{1p}\sum_{l^{\prime}=m}^\infty \xi^{2l^{\prime}+1}
\lambda_{l^{\prime}}\frac{(n+l^{\prime})!(l^{\prime}+l)! }
{(l^{\prime}-m)!(l^{\prime}+m)!} +  \ \nonumber  \\
+R_{1p}^2\sum_{l^{\prime}=m}^\infty \sum_{l^{\prime \prime}=m}^\infty
\xi^{2l^{\prime}+1}\xi^{2l^{\prime \prime}+
1}
\lambda_{l^{\prime}}\lambda_{l^{\prime \prime}} \frac{(n+l^{\prime})!(l^{\prime}+l)! }
{(l^{\prime}-m)!(l^{\prime}+m)!}\frac{(l^{\prime} +l^{\prime \prime})!
(l^{\prime \prime}+l)! }
{(l^{\prime \prime}-m)!(l^{\prime \prime}+m)!} + \cdots
\end{eqnarray}
Using (\ref{nsix}) and (\ref{nseven}) formula (\ref{nnone}) can be significantly
simplified
\begin{equation}
D(\mathbf{r}, \mathbf{r}^{\prime})=\sum_{m=0}^\infty \sum_{l=m}^\infty
 C_{l}^m(\mathbf{r}^{\prime})\left(\frac 1{r^{l+1}}
+\frac{r^l}{A_l}\right)
 P_l^m(\cos \theta)
\Bigg\}\epsilon_m\cos m(\phi - \phi^{\prime}) \label{nntwo}
\end{equation}
Using (\ref{nntwo}) in (\ref{aone}) we obtain
\begin{eqnarray}
 \left \langle \phi (\mathbf{r})  \phi^{*} (\mathbf{r}^{\prime}) \right\rangle_{\omega}= \frac {\hbar}{\pi}\left(n_1(\omega)-n_2(\omega)\right)
\sum_{m=0}^\infty \sum_{n=m}^\infty
\Bigg\{\frac{(n+m)!}{(n-m)!}\frac { C_{n}^m(\mathbf{r})
 C_{n}^{*m}(\mathbf{r}^{\prime})}{R^{2n+1}}            \nonumber  \\
 \times \frac{\mathrm{Im}\lambda_n}{\mid 
\lambda_n \mid^2} \cos m(\phi - \phi^{\prime})\Bigg\} 
 -4\pi A(T_1)\mathrm{Im}D(\mathbf{r}, \mathbf{r}^
{\prime }) \label{atwo}
\end{eqnarray}
Using (\ref{atwo}) in (\ref{twelve}) for the heat transfer between a sphere and
a flat surface we obtain
\begin{eqnarray}
S =\frac{\hbar }{\pi ^2}\int_0^\infty \mathrm{d}\omega \omega \left(n_1(\omega)-
n_2(\omega)\right)
\sum_{m=0}^\infty \sum_{n=m}^\infty\sum_{l=m}^\infty \frac{\xi^{n+l+1}}{(n-m)!
(l-m)!}  \nonumber  \\
\left( (n-m)!\mathrm{Im}\left( \tilde{x}_{nn}^m
R_{1p}^{*}\right)\delta_{nl} - \frac{\xi^{n+l+1}\mid R_{1p}\mid^2}
{(n+l)!(l+m)!}  \mid \tilde{x}_{ln}^m \mid^2 
\mathrm{Im}\lambda_l\right)\mathrm{Im}\lambda_n  \label{athree}
\end{eqnarray}
The above formalism  gives, in principle, an exact solution of the problem 
in the non-retarded limit. However, for  $d<<R$  extensive
 numerical
calculation are necessary, because in this case the series
converge slowly. The numerical results
 will be presented elsewhere. In the present publication
we only present an approximate solution of the problem. 

Using image theorem for $\varepsilon>>1$ and  for the points $\mathbf{r}$
and
 $\mathbf{r}^{\prime}$ close to the surface of the sphere, in the first
approximation in the  expansion of the electrostatic potential in the sum
of the potentials created by the image charges we can write the potential
in the form
\begin{eqnarray}
D(\mathbf{r}, \mathbf{r}^{\prime}) = \frac 1{\mid \mathbf{r}-\mathbf{r}^{\prime}\mid}
-R_{1p}\frac 1{\mid \mathbf{r}-\tilde{\mathbf{r}}^{\prime}\mid}- \nonumber  \\
- R_{2p}\frac 1{\mid \mathbf{r}-\mathbf{r}_i^{\prime}\mid} +
R_{1p}R_{2p}\frac 1{\mid \mathbf{r}-\tilde{\mathbf{r}}_i^{\prime}\mid }, \label{eone}
\end{eqnarray}
where 
\[
\mathbf{r}_i^{\prime} = \frac{2R-r^{\prime}}{r^{\prime}}\left(x^{\prime},
 y^{\prime}, z^{\prime}\right)
\]
$\tilde{\mathbf{r}}^{\prime} = (x^{\prime},
 y^{\prime}, -h -z^{\prime})$ and $\tilde{\mathbf{r}}_i^{\prime} = (x_i^{\prime},
 y_i^{\prime}, -h -z_i^{\prime})$. The value of the surface integral
(\ref{aone}) does not change if we  assume that the potential has the form
(\ref{eone}) in all space outside a sphere. Thus  using  
the Green's theorem  we can convert the surface
integral to a volume integral over all space outside a sphere.  
 This volume integral can be easily calculated using 
the fact that outside a sphere the potential $D(\mathbf{r}, \mathbf{r}^{\prime})$   
obeys the Poisson's equation with the point charges located at $\mathbf{r} =
\mathbf{r}^{\prime}, \mathbf{r} =  \tilde{\mathbf{r}}^{\prime}$, and
$\mathbf{r}= \tilde{\mathbf{r}}_i^{\prime}$  . Performing the  
calculation gives: 
\begin{eqnarray} 
\mathrm{Im}\left \langle \phi (\mathbf{r})  \phi^{*} (\mathbf{r}^{\prime}) 
\right\rangle_{\omega}
= \frac {\hbar}{2\pi}\left(n_1(\omega)-n_2(\omega)\right)\mathrm{Re}
 \Bigg\{ R_{1p}R_{2p}^{*}\frac 1{\mid \mathbf{r}_i^{\prime} - 
\tilde{\mathbf{r}}\mid} -   \nonumber  \\
 R_{1p}^{*}R_{2p}\frac 1{\mid \tilde{\mathbf{r}}^{\prime} - 
\mathbf{r}_i\mid} +
R_{1p}R_{2p}\frac 1{\mid \mathbf{r}^{\prime} -
\tilde{\mathbf{r}}_i\mid} - 
R_{1p}^{*}R_{2p}^{*}\frac 1{\mid \tilde{\mathbf{r}}_i^{\prime} - 
\tilde{\mathbf{r}}\mid}\Bigg\} \label{cone}
\end{eqnarray}
Using (\ref{cone}) in (\ref{twelve}) for the heat transfer between a sphere and
a flat surface we obtain
\begin{eqnarray}
S =\frac{\hbar }{2\pi}\int_0^\infty \mathrm{d}\omega \omega \left(n_1(\omega)-
n_2(\omega)\right)
\mathrm{Im}R_{1p}(\omega )\mathrm{Im}
R_{2p}(\omega ) \nonumber  \\
\times \int_0^\pi \mathrm{d} \theta \frac{\cos^3\theta\sin\theta}
{\left[(2\xi)^{-1}-\cos\theta\right]^2} \approx \frac{\pi}
{30}\frac{R}{d}\sigma_1^{-1}\sigma_2^{-1}\hbar^{-3}k_B^4 (T_1^4-T_2^4), \label{ctwo}
\end{eqnarray}
where $\sigma=\omega_p^2\tau/4\pi$. 
If we assume that in an accordance with (\ref{fifty}) every the elementary volume of 
the sphere gives the contribution to the heat transfer
\begin{equation}
\mathrm{d}S_{pz} \approx \frac{3 k^4_B }{160(R+d-R\cos\theta)^3\hbar^3}
\sigma_1^{-1}\sigma_2^{-1}
(T_1^4-T_2^4)\mathrm{d}V, \label{cthree}
\end{equation}
then after integration of (\ref{cthree}) over the volume of the sphere
we obtain the result of the same order magnitude as (\ref{ctwo}).
Because for $ d<R<d_W \mid \varepsilon\mid^{-1/2}$ in the 
accordance with (\ref{fiftyone}) the $s$-contribution
of small particle does not depend on the separation $d$
formula  (\ref{fiftyone}) is valid also for small separation $d$.
From the comparison (\ref{ctwo}) and (\ref{fiftyone}) we get that
for the sphere close to the surface $s$-contribution dominates
for 
\[
d>10^{-3}(d_Wc/\omega_p)^{3/2}R^{-2}(\omega_p\tau)^{-3/2.}
\]
For ``normal'' metal at room temperature $d>10\,\mathrm{\AA}$.

\section{Heating of a particle by an STM tip}
Let us now consider the heat transfer between an STM tip, which we again model
by a spherical particle with radius $R_2$ and the  polarizability  $\alpha_2(\omega)$,
 and a spherical  particle with radius $R_1$ and the  polarizability  $\alpha_(\omega)$
 located on a surface. We   consider the case of large
separation $d>>R_1, R_2$, and neglect by influence of the substrate on the
heat transfer. At large distances the thermal electromagnetic field
radiated by particle $1$ can be considered as the radiative electromagnetic
field of a fluctuating point dipole $\mathbf{p}^f$ with ensemble average
\begin{equation}
\left\langle p_i^fp_k^{f*}\right\rangle=A(T_1)\mathrm{Im}\alpha_1(\omega)\delta_{ik}
\end{equation}
The electric field of this point dipole is given by
\begin{equation}
\mathbf{E}= [3\mathbf{n}(\mathbf{n\cdot p}^f)-\mathbf{p}^f]\left(\frac 1{r^3}-
\frac{\mathrm{i}k}{r^2}\right)e^{\mathrm{i}kr}-k^2[\mathbf{n}(\mathbf{n\cdot p}^f)-\mathbf{p}^f]
\frac{e^{\mathrm{i}kr}}{r}, \label{fiftytwo}
\end{equation}
where $k=\omega/c$ and where  $\mathbf{n}$ is a unit vector along the axis connecting
the two particle. 
The rate at which a particle ${\bf1}$ does work on a particle ${\bf2}$ is determined by
\begin{eqnarray}
P=4 \int_0^\infty\mathrm{d}\omega\omega\mathrm{Im}\alpha_2(\omega)
\left\langle E_iE_i^{*}\right\rangle_{\omega}
\\
\left\langle E_iE_i^{*}\right\rangle_{\omega}=A(T_1,\omega)\mathrm{Im}\alpha_1(\omega)
\left(\frac 6{d^6}+\frac{2k^2}{d^4}+\frac{2k^4}{d^2}\right)
\end{eqnarray}
After absorption in the particle $2$ this work is converted into heat. In the same
manner we can calculate the rate of cooling of particle $2$ using the same formula
by reciprocity. Thus the total heat transfer between the particles will determined by
\begin{eqnarray} 
S=\frac{\hbar}{(2\pi)^2}\int_0^\infty \mathrm{d} \omega \omega [n_1(\omega)-n_2(\omega)] 
   \mathrm{Im}\alpha_1(\omega)\mathrm{Im}\alpha_2(\omega)  \nonumber   \\
\times\left(\frac 6{d^6}+\frac{2k^2}{d^4}+\frac{2k^4}{d^2}\right)
\end{eqnarray}
Using for the  polarizabilities of particles expression (\ref{sixty}) we obtain

\begin{eqnarray}
S \approx 10^{-1}\hbar^{-3}k_B^4T_1^4R_1^3R_2^3 \sigma_1^{-1} \sigma_2^{-1} \left(\frac 6{d^6}
+2\pi^2\frac 1{d_{1W}^2d^4}+8\pi^4\frac 1{d_{1W}^4d^2}\right) \nonumber  \\
 - [T_1\rightarrow T_2],
\end{eqnarray} 
where $d_{iW}=\hbar c/k_BT_i$.

\section{Relation between heat transfer and friction}
The heat transfer studied above is closely related to the frictional stress
between bodies in relative motion, separated by a vacuum gap. In the last
years this ``vacuum'' friction has attracted a great attention in the
connection with the development of the scanning probe technique
\cite{Volokitin, Persson and Zhang, Persson1, Dorofeev, Dedkov, Pendry1}. In
\cite{Volokitin} we show that the frictional stress between the bodies
having flat parallel surfaces separated by a distance $d$ and moving
with velocity $V$ relative to each other is determined by formula which is very
similar to (\ref{seventyone}):
\begin{eqnarray}
\sigma &=&\frac { \hbar V}{16\pi ^3}\int_0^\infty \mathrm{d}\omega
\int_{q<\omega /c }\mathrm{d}^2q q^2 \nonumber \\
&&\times\left[
\frac{(1-\mid R_{1p}(\omega )\mid ^2)(1-\mid R_{2p}(\omega)\mid ^2)%
}{\mid 1-%
\mathrm{e}^{2\mathrm{i}pd}R_{1p}(\omega )R_{2p}(\omega )\mid ^2}\right]\left(-
\frac {\partial n(\omega )}{\partial \omega}\right)
\nonumber \\
&&  +\frac {\hbar V}{4\pi
^3}\int_0^\infty \mathrm{d}\omega \int_{q>\omega /c}%
\mathrm{d}^2q q^2\mathrm{e}^{-2\mid p\mid d}  \nonumber \\
&&\times  \frac{\mathrm{Im}R_{1p}(\omega )\mathrm{Im}%
R_{2p}(\omega )}{\mid 1-\mathrm{e}^{-2\mid p\mid d}R_{1p}(\omega
)R_{2p}(\omega)\mid ^2}\left(-
\frac {\partial n(\omega )}{\partial \omega}\right)
\nonumber \\
&&\left. +\left[ p\rightarrow s\right]  \label{cfive} \right.
\end{eqnarray}
For the distances $d<<d_W$ this formula is reduced to:
\begin{eqnarray}
\sigma =\frac{\hbar }{2\pi ^2}\int_0^\infty \mathrm{d}\omega
\int_0^\infty \mathrm{d}qq^3\mathrm{e}
^{-2qd}   \nonumber  \\
\times \left\{ \frac{\mathrm{Im}R_{1p}(\omega )\mathrm{Im}
R_{2p}(\omega )}{\mid 1-\mathrm{e}^{-2qd}R_{1p}(\omega )R_{2p}(\omega )\mid
^2}\left(-
\frac {\partial n(\omega )}{\partial \omega}\right) +[p\rightarrow s]\right\},
  \label{csix}
\end{eqnarray}
which is very similar to the corresponding heat-transfer formula (\ref{fourtyeight}).
For $d<d_W\varepsilon^{-1/2}$, where $\varepsilon(\omega)$ is taken at
the characteristic frequency $\sim k_BT/\hbar$, the $p$- wave
contribution is given by \cite{Volokitin}
\begin{equation}
\sigma_p \approx 0.3\left( k_BT/\hbar\omega_p\right)^2\frac 1 {(\omega_p\tau)^2}
\frac {\hbar V}{d^4}  \label{cseven},
\end{equation}
and for $d_W\varepsilon^{-1/2}<d<d_W$:
\begin{equation}
\sigma_p \approx\frac 9{2\pi^2}\frac{\hbar V}{d^2d_W^2}\frac{ k_BT}{\hbar\omega_p}
\frac {1} {\omega_p \tau}.  \label{done}
\end{equation}
For $d<d_W\varepsilon^{-1/2}$, the $s$-wave contribution $\sigma_s$ becomes independent
 of $d$:
\begin{equation}
\sigma_s\approx 0.5 \pi^{-2}\hbar^{-1}c^{-4}k_B^2T^2\tau^2\omega_p^4V. \label{done}
\end{equation}
For $d>d_W\varepsilon^{-1/2}$ we have:
\begin{equation}
\sigma_s\approx3.88 \pi^{-2}d^{-2}c^{-2}k_B^2T\tau\omega_p^2V.
\end{equation}
From the comparison (\ref{cseven}) and  (\ref{done}) we find that $\sigma_s
>\sigma_p$ for $d>c/\omega_p^2\tau$. For typical metal at room temperature
this corresponds to $d>1\,
\mathrm {\AA}$. This is in drastic constant to the (conservative) van der Waals
interaction, where the retardation effects become important only for
 $ d>c/\omega_p $ \cite{Lifshitz}.
Finally we carry out the transition to frictional stress between a particle with the radius
$R<<d$ and semi-infinite body in (\ref{csix}). To do this, we assume as in Sec.4 that the body
$\mathbf{2}$ is sufficiently rarefied, i.e., that the difference $\varepsilon_2-1$ is small.
Keeping only the first non-vanishing terms in the expansion of the integrand of (\ref{csix})
in powers of these difference, we get formula similar to (\ref{fourtynine}): 
\begin{eqnarray}
\sigma =\frac{\hbar V  }{\pi}\int_0^\infty \mathrm{d}\omega \left(-\frac {\partial n(\omega )}{\partial 
\omega}
\right)
\int_0^\infty \mathrm{d}qq^4\mathrm{e}
^{-2qd}  \nonumber \\
\times \left\{ 2\mathrm{Im}R_{1p}(\omega )\mathrm{Im}
\alpha_{2}(\omega )
+\left(\frac{\omega}{cq}\right)^2\mathrm{Im}R_{1s}(\omega )\mathrm{Im}
\alpha_{2}(\omega  \right\}
  \label{dtwo}
\end{eqnarray}
In the limit $d<\mid \varepsilon \mid^{-1/2}d_W$
 the reflection factor of $p$-wave is determined by (\ref{dthree}) and in (\ref{dtwo}) the $p$-wave
contribution  is reduced to the formula, which was obtained by Tomassone and Widom \cite{Tomassone}.
 For the spherical particle in this limit we get
\begin{equation}
\sigma_p\approx \frac {3}{16\pi^2}\frac {\hbar V}{d^5}\left(\frac{k_BT}{\hbar}\right)^2
\sigma^{-1}_1\sigma^{-1}_2 R^3, \label{fone}
\end{equation} 
and for $\mid \varepsilon \mid^{-1/2}d_W<d<d_W$ we get
\begin{equation}
\sigma_p\approx 14\pi^{-5/2}\frac {\hbar V}{d^4}\left(\frac{k_BT}{\hbar}\right)^{5/2}
\sigma^{-1/2}_1\sigma^{-1}_2c^{-1} R^3. \label{ftwo}
\end{equation} 
For  $d<\mid \varepsilon \mid^{-1/2}d_W$, $\sigma_s$ is independent of $d$:
\begin{equation}
\sigma_s\approx3.3\cdot10^2\pi^{-1/2}\hbar V \sigma^{3/2}_1\sigma^{-1}_2c^{-5}R^3
\left(\frac{k_BT}{\hbar}\right)^{9/2}, \label{fthree}
\end{equation}
and for $\mid \varepsilon \mid^{-1/2}d_W<d<d_W$ we get:
\begin{equation} 
\sigma_s\approx3.54\pi^{-3/2}\frac {\hbar V}{d^4}\left(\frac{k_BT}{\hbar}\right)^{5/2}
\sigma^{-1/2}_1\sigma^{-1}_2c^{-1} R^3.  \label{ffour}
\end{equation}
From the comparison (\ref{fone}) and (\ref{fthree}) we get that $\sigma_s>\sigma_p$
for $0.1(cd_W/\sigma_1)^{1/2}<d<\mid \varepsilon \mid^{-1/2}d_W$. For normal metal at room 
temperature $10^2<d<10^3\,\mathrm{\AA}$. For  $\mid \varepsilon \mid^{-1/2}d_W<d<d_W$
$\sigma_s\approx \sigma_p$.

To estimate $\sigma$ for $R>>d$ we use the same approach as in Sec. 4. We define the
frictional stress between the elementary volume $\mathrm{d}V$ and the semi-infinite body 
as $\mathrm{d}\sigma=(3\sigma/4\pi R^3)\mathrm{d}V$, where $\sigma$ is given by an expression
for a spherical particle for $R<<d$. After integration over the volume of the sphere
for $d<<R<\mid \varepsilon \mid^{-1/2}d_W$ we get:
\begin{equation}
 \sigma_p\approx 4\cdot10^{-5 }\frac {\hbar V}{d^3}\left(\frac{k_BT}{\hbar}\right)^2
\sigma^{-1}_1\sigma^{-1}_2 R. \label{ffive}
\end{equation}
Because in this limit $\sigma_s$ does not depend on $d$, it is still determined
by (\ref{ffour}). From (\ref{ffour}) and (\ref{ffive}) we get that $\sigma_s>\sigma_p$
for
\[
d>2.4\cdot10^4(cd_W\sigma^{-1}_1)^{5/6}R^{-2/3}.
\]
 For ``normal'' metal at room temperature and $R\sim10^3\,\mathrm{\AA}$, $d>10\,\mathrm{\AA}$.

Recently, Dorofeev \textit{et.al}. \cite{ Dorofeev} have observed Brownian motion of a
small metal particle connected by a spring to a holder, and located in ultrahigh vacuum in the vicinity of a gold surface. It was observed that the particle performed a stochastic oscillatory motion increased as the particle approach the gold surface. It was suggested that this increased damping is due to the coupling to the fluctuating electromagnetic field. From (\ref{fthree})
and (\ref{ffive}) we can estimate the damping constant $\gamma/m=\sigma/mV$. For $d\sim 10^2
\,\mathrm{\AA}, R\sim 10^3\,\mathrm{\AA}, m\sim10^{-11}-10^{-13}\,kg$, and for ``normal''
metal at room temperature we get $\gamma_p/m\sim 10^{-17}\,s^{-1}$ and 
$\gamma_s/m\sim 10^{-13}\,s^{-1}$. However in the experiment \cite{ Dorofeev}
 $\gamma/m\sim10^2\,s^{-1}$. Thus the contribution of a fluctuating electromagnetic field
to the damping constant can not explain the observed experimental date. This result is in
an agreement with
our earlier conclusion \cite{Persson1}.

The fluctuating electromagnetic field is an origin of the frictional drag, observed
between parallel two-dimensional electron systems \cite{Gramila}. In the frictional
drag experiments a current is drawn in the first layer, while the second layer is an open 
circuit. Thus no dc - current can flow in the second layer, but an induced electric field occur
that opposes the ``drag force'' from the first layer. Recently we used the theory 
of a fluctuating electromagnetic field to calculate frictional drag force between
 two-dimensional electron systems \cite{Volokitin1}. For frictional drag stress we
found formula, which is very similar to (\ref{cfive}). We found that for modulation-doped semiconductor quantum wells retardation effects are not important under typical experimental 
conditions, supporting earlier calculations where retardation effects always have been
neglected \cite{Gramila, Pogrebinskii}. A striking new result we found is that 
for systems with high 2D-electron density, e.g., thin metallic films, retardation effects becomes crucial and in fact dominate the frictional shear stress $\sigma$.

\section{Summary and conclusion}

We have developed a general formalism for the calculation of the radiative
heat transfer between two bodies of arbitrary shape and properties. The formalism
has been applied to three different cases, namely the heat transfer between (a) two flat surfaces, (b) a spherical particle and a flat surface, and (c) between
two spherical particles. For  two flat solid surfaces we have presented
numerical results for several cases of practical importance, namely for 
two ``normal'' (high conductivity) metals (silver), 
two (high resistivity) semi-metals 
and two silver
metals coated by thin layers ($10 \ {\rm \AA}$) of high resistivity material.
For high-resistivity metals the $p$-wave contribution dominates, but for ``normal''
(high conductivity) metals we found the remarkable result that the $s$-wave
contribution dominates even for short separation between the solids.
We have pointed out the close relation ship between the radiative
heat transfer between two solids and the vacuum friction [..] which occur when one
of the solids slide relative to the other solid. The formalism developed in
this paper may can be generalized to treat the vacuum friction between bodies with
curved surfaces.

\vskip 0.5cm
\textbf{Acknowledgment }

A.I.V acknowledges financial support from DFG  and thank R.O.Jones for many important 
 references and for help in preparation of the manuscript. B.N.J. P acknowledges
financial support from BMBF.
\vskip 1cm

 FIGURE CAPTIONS

Fig. 1 (a) The heat transfer flux between two semi-infinite silver bodies, one
at
temperature $T_1 = 273 \ {\rm K}$ and another at $T_2 = 0 \ {\rm K}$. (b) The
same as
(a) except that we have reduced the Drude electron relaxation time $\tau$ for
solid 1
from
a value corresponding to a mean free path $v_F \tau = l = 560 \ {\rm \AA}$ to
$20 \ {\rm \AA}$. (c) The same as (a) except that we have reduced $l$ to $3.4 \
{\rm \AA}$.

Fig. 2 The thermal flux as a function of the separation $d$ between the
surfaces.
One body is at zero temperature and the other at $T=273K$. With $(\hbar
/\tau)/k_BT
=120$ and $\hbar \omega_p /k_BT = 15.5$.

Fig. 3 The thermal flux as a function of the resistivity of the solids.
The solid surfaces are separated by $d = 10 \ {\rm \AA}$ and $\hbar \omega_p >>
k_BT$.
The heat flux for other separations can be obtained using scaling $\sim 1/d^2$
which
holds for high-resistivity materials.

Fig. 4. The heat flux between two semi-infinite silver bodies coated with
$10 \, {\rm \AA}$ high resistivity ($\rho = 0.14 \ {\rm \Omega cm}$) material.
Also shown is the heat flux between two silver bodies, and two high-resistivity
bodies.
One body is at zero temperature and the other at $T=273K$.
(a) and (b) shows the $p$ and $s$-contributions, respectively.

Fig. 5. Spherical particle (origin $O$) above a flat surface and its
``image'' (origin $O^{\prime}$).


\begin{thebibliography}{99}

\bibitem{Van Hove} D.Polder and M.Van Hove, Phys. Rev. B \textbf{4}, 3303, (1971).

\bibitem{Pendry} J.B.Pendry, J.Phys.:Condens.Matter \textbf{11}, 6621 (1999).
\bibitem{Majumdar} A.Majumdar, Ann.Rev:Mater.Sci.  \textbf{29}, 505 (1999).

\bibitem{Rytov}  S. M. Rytov, \textit{Theory of Electrical Fluctuation and
Thermal Radiation} (Academy of Science of USSR Publishing, Moscow, 1953).


\bibitem{Lifshitz}  E. M. Lifshitz, Zh.Eksp. Teor. Fiz. \textbf{29}, 94
(1955) [Sov. Phys.JETP \textbf{2}, 73 (1956).

\bibitem{Volokitin}  A. I. Volokitin and B. N. J. Persson, J.Phys.: Condens.
Matter \textbf{11}, 345 (1999); \textit{ibid. }Phys.Low-Dim.Struct. \textbf{%
7/8, }17 \textbf{\ }(1998).

\bibitem{Abrikosov}  A.A.Abrikosov, L.P.Gor'kov and I.Ye.Dzyaloshinskii,
\textit{Quantum Field Theoretical Methods in Statistical Physics}
(Pergamon Press Ltd., 1965



\bibitem{Persson and Zhang}  B. N. J. Persson and Z. Zhang, Phys. Rev. B
\textbf{57}, 7327 (1998).

\bibitem{Landau}  L. D. Landau and E. M. Lifshitz,\textit{\ Electrodynamics
of Continuous Media} (Pergamon Press Ltd., 1975).



\bibitem{Nozawa} R.Nozawa, J.Math.Phys. \textbf{7}, 1841 (1966)

\bibitem{Persson1} B.N.J.Persson and A.I.Volokitin, Phys. Rev. Lett. \textbf{84},
3504 (2000).

\bibitem{Dorofeev} I.Dorofeev, H.Fuchs, G.Wenning, and B.Gotsman, Phys. Rev. Lett.
\textbf{83}, 2402 (1999).

\bibitem{Dedkov} G.V.Dedkov and A.A.Kyasov, Phys.Lett. A \textbf{259}, 38, (1999). 

\bibitem{Pendry1} J.B.Pendry,  J.Phys.:Condens.Matter  \textbf{9}, 10301, (1997).

\bibitem{Tomassone} M.S.Tomassone and A.Widom, Phys.Rev. B  \textbf{56}, 4938,
  (1997)
\bibitem{Gramila} T.J.Gramila, J.P.Eisenstein, A.H.Macdonald, L.N.Pfeiffer, and
K.W.West, Phys.Rev. \textbf{66}, 1216 (1991); Surface.Sci.,  \textbf{263},
 446, (1992); Phys.Rev. B \textbf{47},12 957, (1993); Physica B \textbf{197},
 442, (1994).

 
\bibitem{Volokitin1} A.I.Volokitin and B.N.J.Persson, (to be published).

\bibitem{Pogrebinskii} M.B.Pogrebinskii, Fiz.Tekh.Poluprovodn. \textbf{11}, 637,
 (1977) [Sov.Phys.Semicond. \textbf{11}, 372 (1977)]; P.J.Rice, Physica B\&C \textbf{117},
 750, (1983).
 


\end{thebibliography}
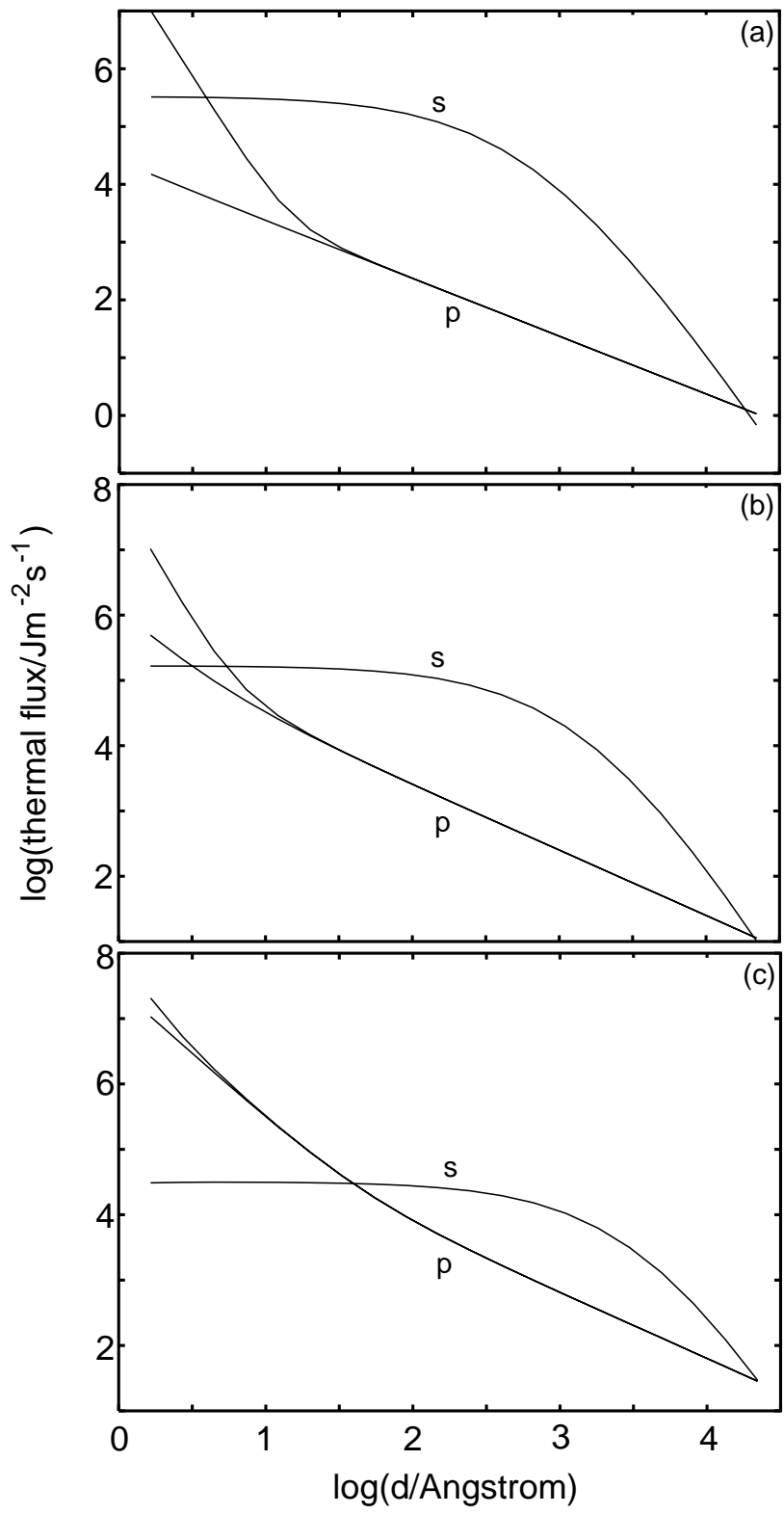
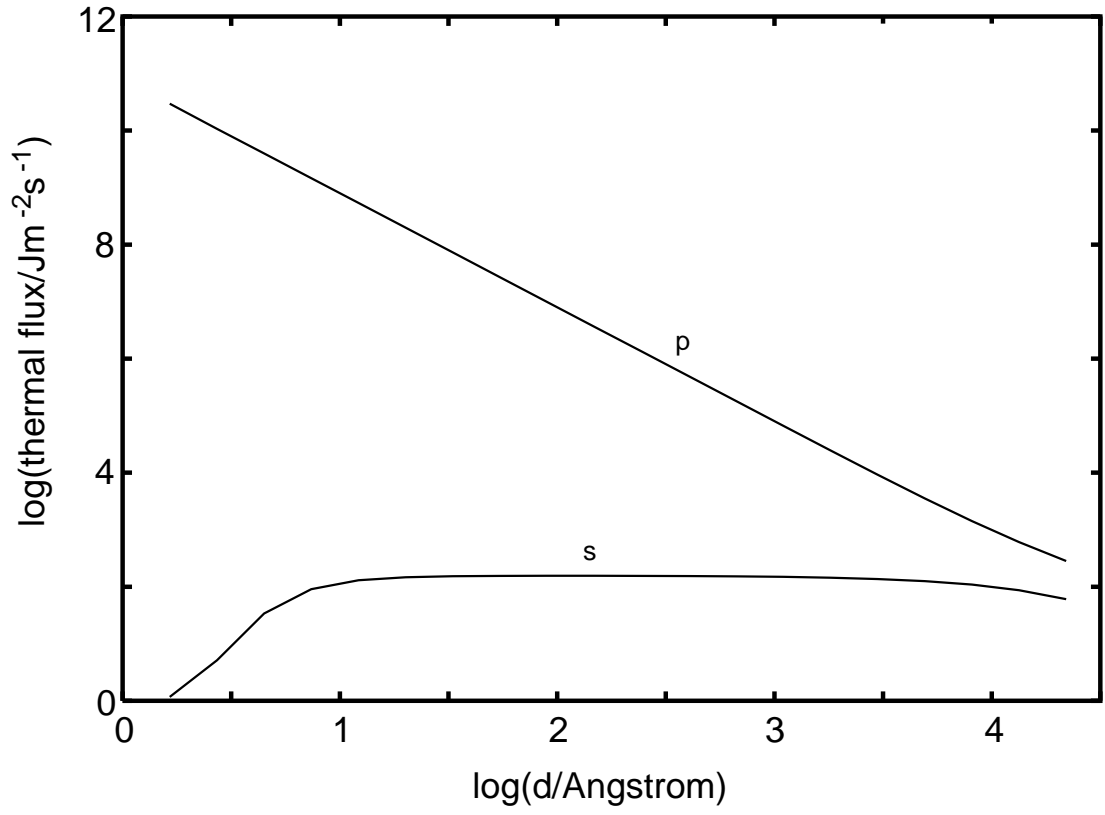
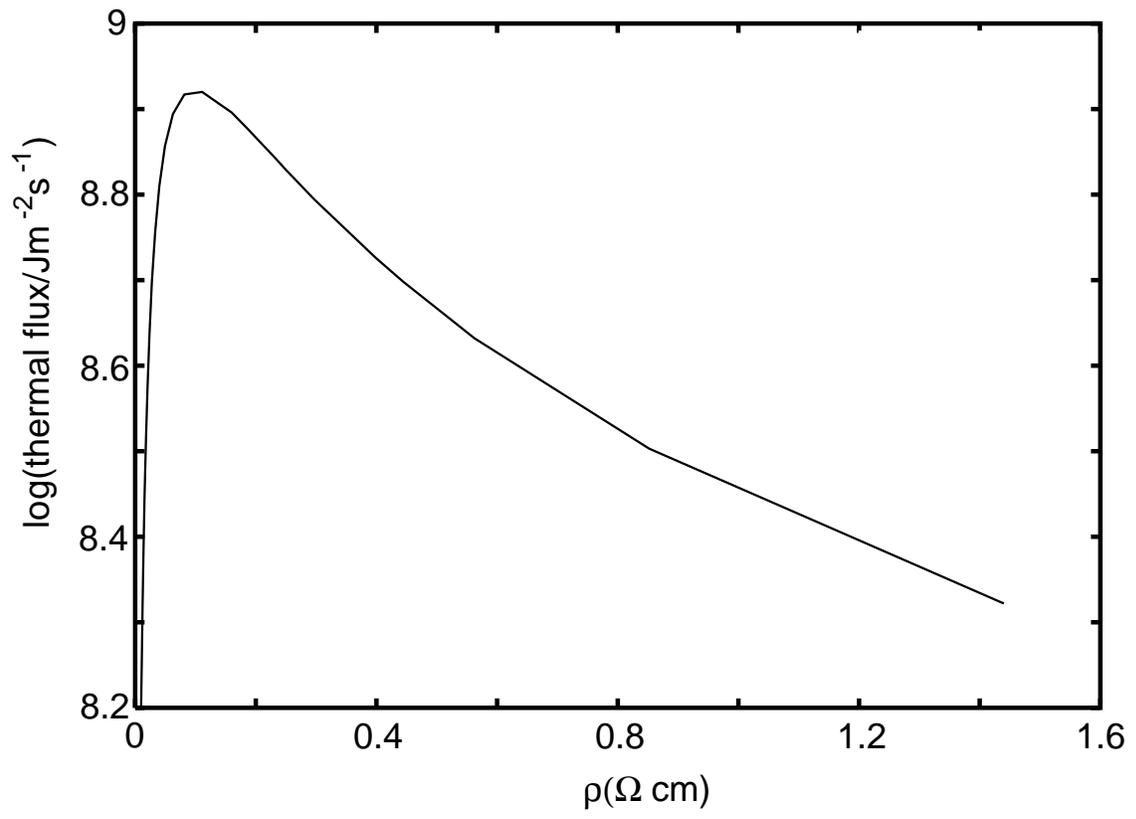
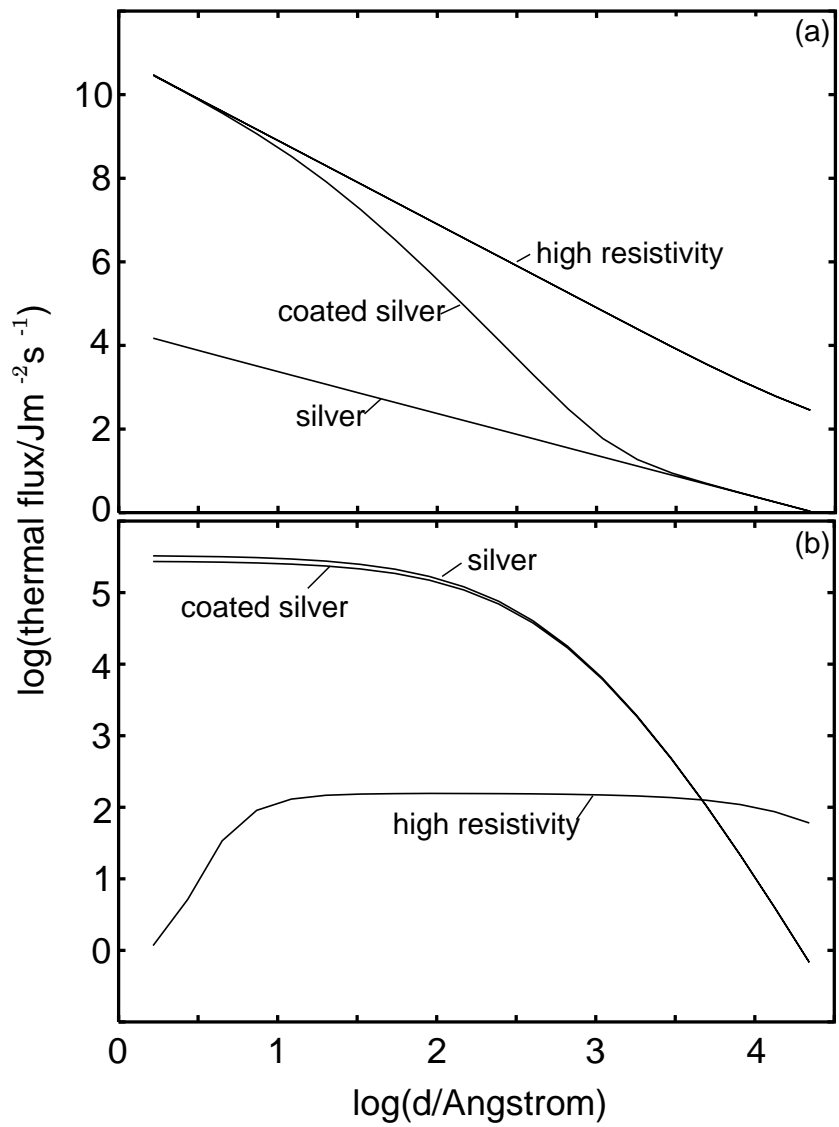
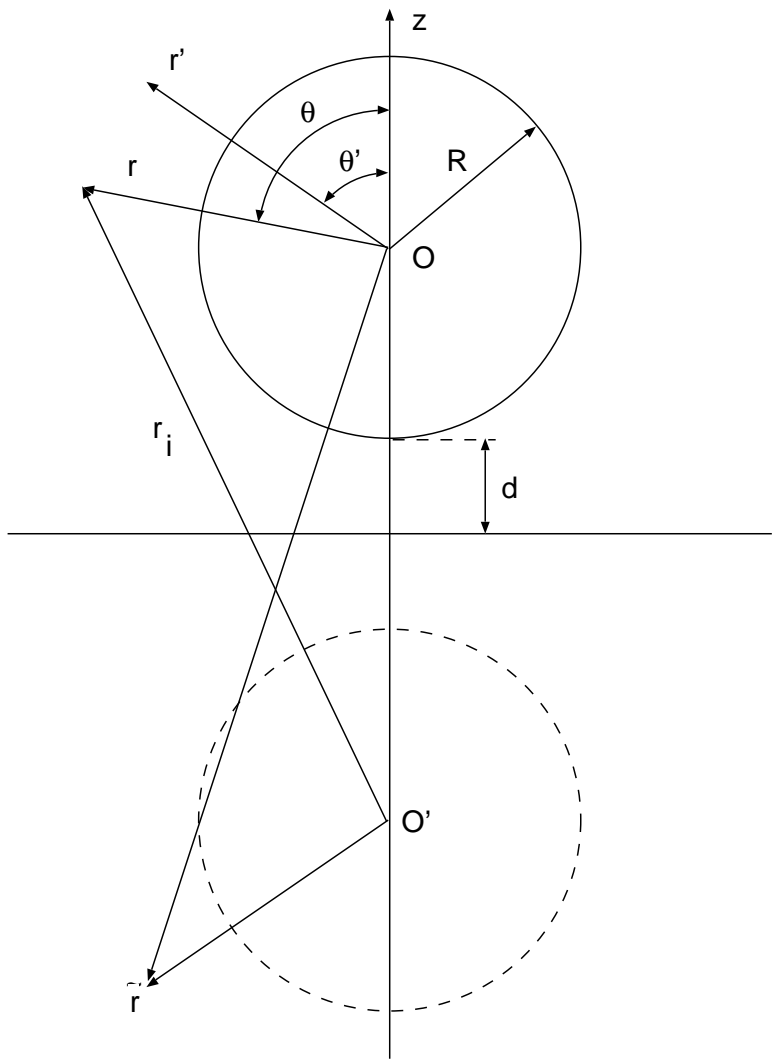
\end{document}